# Chapter 11

# 11 T Dipole for the Dispersion Suppressor Collimators

*M. Karppinen[1], S. Izquierdo Bermudez[1], A. Nobrega[2], H. Prin[1], D. Ramos[1], S. Redaelli[1], F. Savary[1]\*, D. Smekens[1] and A. Zlobin[2]*

[1]CERN, Accelerator & Technology Sector, Geneva, Switzerland
[2]FNAL, Fermi National Accelerator Laboratory, Batavia, USA

## 11    11 T dipole for the dispersion suppressor collimators

### 11.1    Introduction

A pair of 11 T dipoles will replace some of the main dipoles (MB) in the dispersion suppressor (DS) regions of the LHC to create space for additional collimators, which are necessary to cope with beam intensities that are larger than nominal, such as in the High Luminosity LHC (HL-LHC) project [1].

A joint research and development (R&D) programme was initiated in October 2010 at the Fermi National Laboratory (FNAL) in the US, and in the middle of 2011 at CERN, with the goal of developing the necessary technology for the fabrication of a full-length two-in-one aperture $Nb_3Sn$ dipole prototype suitable for installation in the LHC [2]. After the design, fabrication, and test of a number of short models with a length of 1 m and 2 m, FNAL is now slowing down development, while CERN is gradually ramping up with the fabrication of 2 m long models and the preparation of the tooling for full-length prototypes. The design of the 11 T dipole described in this preliminary design report features the solutions developed in the framework of the CERN programme. Except for the pole loading concept, for the cable insulation scheme and for features specific to full-length magnets, the solutions used at CERN are largely based on the results of the R&D programme conducted at FNAL.

### 11.2    The cryo-assembly

11.2.1    Description

An MB cryo-assembly will be replaced with a string of three independently installed and aligned cryo-assemblies: two of these will each house a 6.252 m long 11 T dipole, referred to below as the MBH, with a bypass cryostat installed between them. The bypass cryostat ensures the continuity of the cryogenic and electrical circuits and comprises cold to warm transitions on the beam lines in order to create a room temperature vacuum sector for the collimator.

Figure 11-1 shows a schematic layout of the string of cryostats composing the 11 T cryo-assembly, which will replace an MB cryostat.

The cryostat for the MBH shall follow the same design and fabrication principles as the other arc cryostats; it shall comply with the static heat loads specified by the Heat Load Working Group [3]. Standard LHC cryostat performance in terms of alignment tolerances and geometrical stability shall be ensured.

---

\* Corresponding author: Frederic.Savary@cern.ch

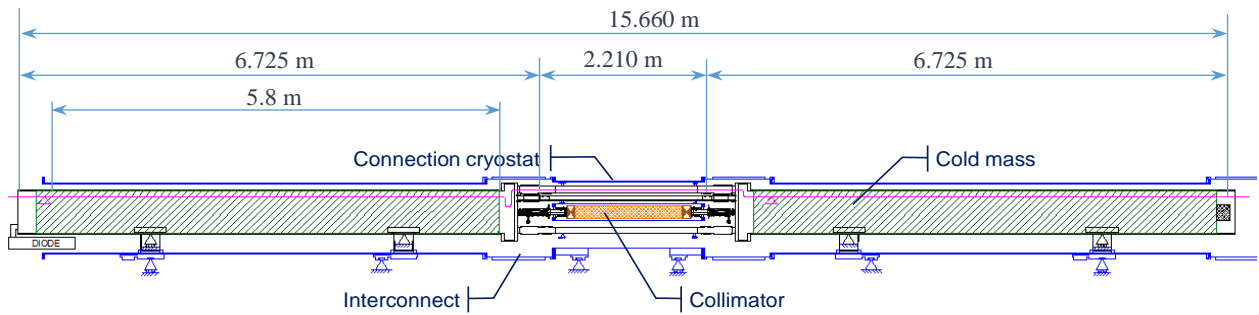

Figure 11-1: Longitudinal section of the 11 T cryo-assembly showing the collimator and the cold-to-warm transitions.

The design of the bypass cryostat shall be compatible with the integration of the collimator and also with the RF-shielded gate valves at the extremity of the cold-to-warm transitions. All cryogenic lines and powering busbars shall have their continuity ensured across the bypass cryostat.

In the present concept, independent installation and alignment of the three cryostats is foreseen. In addition, the TCLD collimator shall be supported directly by the tunnel floor so as not to be affected by deformations of the cryostat vacuum vessels due to alignment or pressure-induced forces.

### 11.2.2 Equipment parameters

The main parameters of the 11 T cryo-assembly are given in Table 11-1. The lengths are, for the present, provisional and may vary depending on the detail design of the MBH and on the details of collimator integration. The dimensions of the cryogenic pipes are equivalent to those of a standard LHC arc continuous cryostat.

Table 11-1: Main parameters of the 11 T cryo-assembly

| Characteristics | Unit | Value |
| --- | --- | --- |
| Total length including interconnects | [mm] | 15 660 |
| Upstream cryostat length between interconnect planes | [mm] | 6 725 |
| Downstream cryostat length between interconnect planes | [mm] | 6 725 |
| Bypass cryostat length between interconnect planes | [mm] | 2 210 |
| Beam line cold bore diameter (inner) | [mm] | 50 |
| Length of room temperature beam vacuum sector measured between cold-to-warm transition flanges | [mm] | 1 550 |

The preliminary design of the 11 T cryo-assembly is based on the following assumptions.

- The interface between the cold beam lines of the MBH cryostats and the beam vacuum sector of the collimator requires sectorization by RF-shielded gate valves.

- As opposed to other collimators in the machine, residual radiation to personnel is assumed to be compatible with the removal and installation of the TCLD collimator without remote handling equipment. Given the integration constraints in the LHC dispersion suppressors, the design of a collimator compatible with remote handling is most likely not achievable.

- Radiation doses on the cryostat throughout the HL-LHC lifetime are compatible with the usage of LHC standard cryostat materials.

- Magnetic shielding is not required on the bypass cryostat. It is assumed that the magnetic field created by the busbar currents will not be detrimental to the accuracy of the TCLD instrumentation and controls.

## 11.3 The 11 T dipole

The design of the MBH is based on the two-in-one concept, i.e. the cold mass comprises two apertures in a common yoke and shell assembly. The MBH cold mass assembly has a length of 6.252 m between the datum planes C and L that are shown on the end covers, see Figure 11-2. The coils have a length of 5.415 m without the outermost end spacers (called saddles), and 5.573 mm with the saddles, see Figure 11-3. A pair of MBHs is needed to produce an integrated field of 119 T m at 11.85 kA, which corresponds to the bending strength of the MB. The MBHs need to be compatible with the LHC lattice and its main systems. They will be connected in series with the MBs.

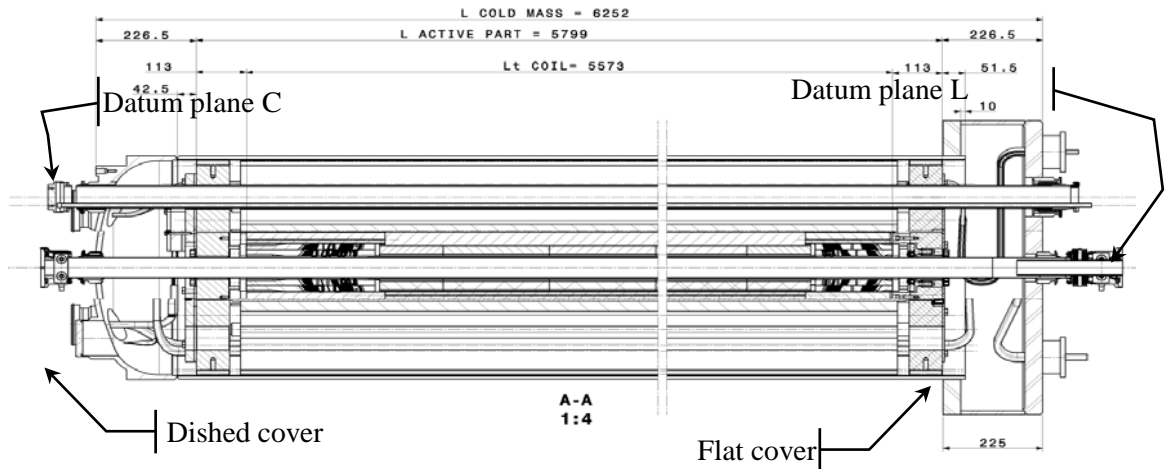

Figure 11-2: Longitudinal section of the cold mass assembly of the MBH

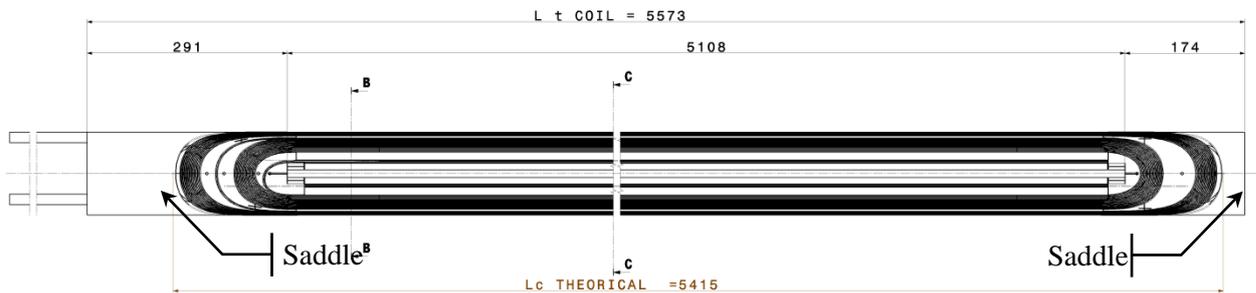

Figure 11-3: Longitudinal section of the coil

### 11.3.1 Description

The coils of the MBH are made of two layers and six blocks with $Nb_3Sn$ keystoned Rutherford-type cable as shown in Figure 11-4 . Each coil comprises 56 turns, with 22 in the inner layer and 34 in the outer layer. There is no splice at the layer jump, i.e. the two layers are wound from the same cable unit length. The cable is made of 40 strands of 0.7 mm diameter. Two manufacturing routes are considered for the strand: the restacked rod (RRP) process and the powder-in-tube (PIT) process.

    The mechanical structure comprises separate austenitic steel collars for each aperture to balance the electromagnetic forces, and a vertically split iron yoke surrounded by a welded stainless steel shrinking cylinder, which contributes to the overall rigidity of the assembly. The axial component of the electromagnetic forces is also transferred to the shrinking cylinder via thick end plates, which are welded to it, and bolts in contact with the saddles. The bolts are screwed into the end plates. The cold mass envelope is closed at the ends by a dished cover on the side of the assembly facing the existing MBs and by a flat cover of a larger diameter on the side of the assembly facing the collimator, see Figure 11-1. A larger diameter is needed on that side to allow the routing of the busbars across the bypass cryostat within the limits of the radial and longitudinal space available.

A cross-section through the MBH is shown in Figure 11-4, without the busbars, heat exchanger tube, support pads, and line N. The key parts of the collared coil are also shown in Figure 11-4.

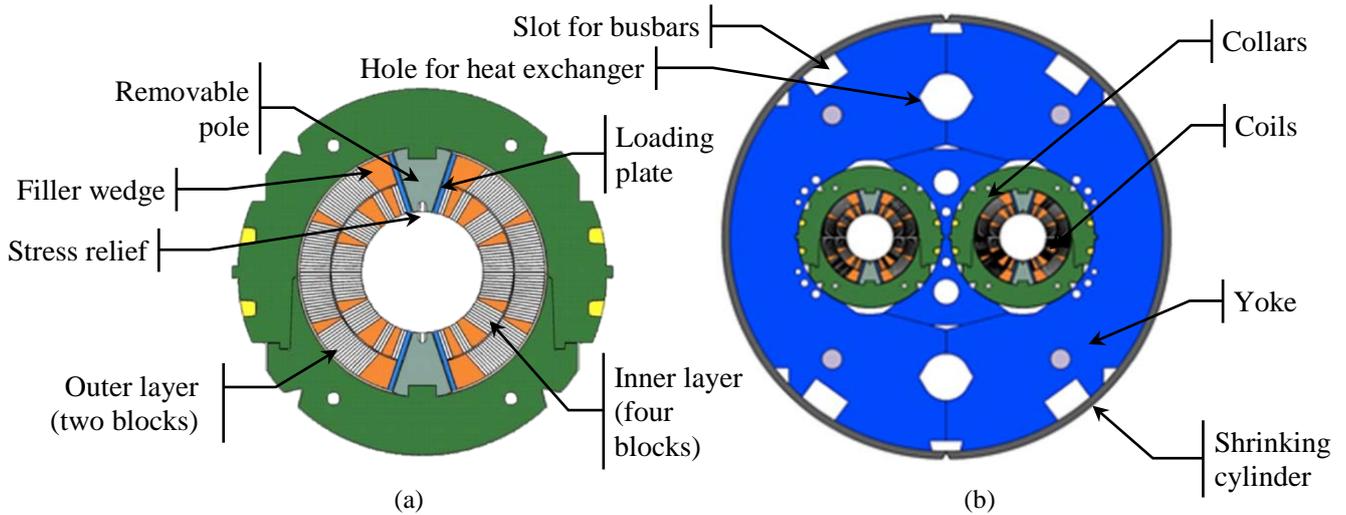

Figure 11-4: Cross-section through (a) the 11 T dipole collared coil; (b) cold mass assembly

To avoid deformation of the beam closed orbit, the integrated transfer function of a pair of MBHs shall be identical to that of the MB. However, this is not possible across the entire range of current during ramping up to nominal, as shown in Figure 11-5. The design is such that a pair of MBHs provides the same integrated field of 119 T m as a standard MB at the nominal current of 11.85 kA.

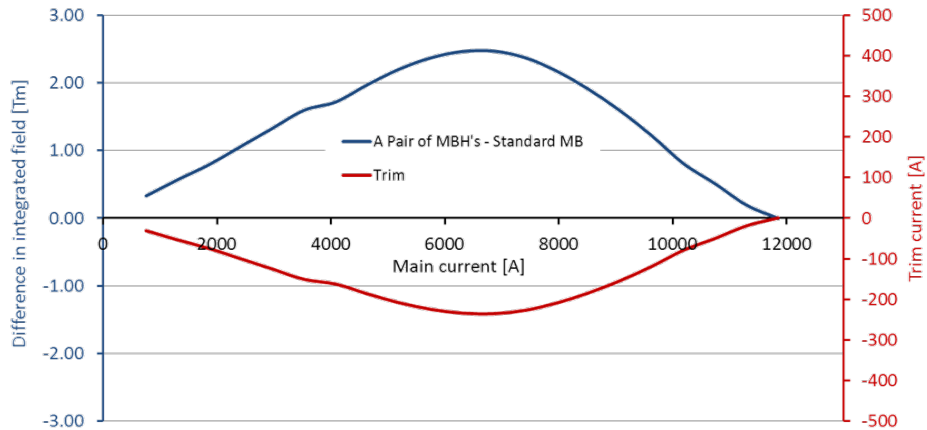

Figure 11-5: The blue line shows the difference in integrated field between a pair of MBHs and an MB, both delivering 119 T m at 11.85 kA. The red line shows the trim current needed to correct the difference at currents below 11.85 kA.

The MBH is stronger at lower currents (it has more turns) with a peak difference in integrated field around 6.5 kA. This can be mitigated by adding a dedicated trim power converter. In the absence of trim current, the resulting orbit distortion could be mitigated by means of the standard orbit correctors in the LHC lattice. However, a fully validated solution including machine protection, reliability, and availability is not currently available (for example, the operation of LHC would be compromised in the case of failure of an orbit corrector). The correction of the transfer function with a trim current is the preferred option as it would allow simpler and more transparent operation.

Unlike the MB, which is curved in the horizontal plane, the MBH will be straight because of the brittleness of $Nb_3Sn$ after reaction. Also, it will be equipped with the same cold bore tube and beam screen as the present curved MB, to facilitate integration. To mitigate for the corresponding reduction of mechanical aperture, the two MBHs of a cryo-assembly will be assembled with an angle of 2.55 mrad relative to each other, and shifted by 0.8 mm towards the centre of the machine.

Depending on the location, the MB to be replaced may be of type A or type B, i.e. it will have both a magnet corrector sextupole (MCS) on the lyre side (i.e. downstream), and a nested magnet corrector decapole/octupole (MCDO) on the connection side (i.e. upstream), respectively, or only an MCS on the lyre side. The MBH, which is on the righthand side of the collimator for an observer looking at the machine from its centre, will have only an MCS on the lyre side. In the current stage of the project, it is also planned to install an MCDO on the connection side of the MBH that is installed on the lefthand side of the collimator. The MCDO may be connected, or not, depending on the installation location of the cryo-assembly.

### 11.3.2 Equipment parameters

A pair of MBHs will provide an integrated field of 119 T m at the nominal operation current of the MBs, 11.85 kA. This corresponds to a nominal magnetic flux density of 11.23 T at the centre of the bore. This goal shall be obtained with a margin of ~20% on the magnet load line [4].

The geometric field quality will be optimized to keep the low-order field errors below 1 unit. The different contributions to the field errors are given in Table 11–2. These include the contributions of the coil ends, cryostat, persistent currents, and eddy currents effects. The field errors shall be confirmed by magnetic measurements. The main parameters of the MBH are listed in Table 11-3.

Table 11-2: Field errors – $R_{ref}$ = 17 mm

|  | Systematic | | | | | Uncertainty | | Random | |
| --- | --- | --- | --- | --- | --- | --- | --- | --- | --- |
| **Normal** | **Geometric** | **Saturation** | **Persistent** | **Injection** | **High field** | **Injection** | **High field** | **Injection** | **High field** |
| 2 | 0.000 | -12.200 | 1.010 | 1.010 | -12.200 | 1.930 | 1.930 | 1.9300 | 1.930 |
| 3 | 7.459 | -0.279 | -1.299 | 6.160 | 7.180 | 1.240 | 1.240 | 1.2400 | 1.240 |
| 4 | 0.000 | -0.400 | 0.070 | 0.070 | -0.400 | 0.600 | 0.600 | 0.6000 | 0.600 |
| 5 | -0.014 | 0.514 | 6.594 | 6.580 | 0.500 | 0.310 | 0.310 | 0.3100 | 0.310 |
| 6 | 0.000 | -0.020 | 0.000 | 0.000 | -0.020 | 0.180 | 0.180 | 0.1800 | 0.180 |
| 7 | -0.093 | 0.062 | -0.688 | -0.780 | -0.030 | 0.110 | 0.110 | 0.1100 | 0.110 |
| 8 | 0.000 | 0.000 | 0.000 | 0.000 | 0.000 | 0.060 | 0.060 | 0.0600 | 0.060 |
| 9 | 0.912 | 0.028 | 1.024 | 1.936 | 0.940 | 0.030 | 0.030 | 0.0300 | 0.030 |
| 10 | 0.000 | 0.000 | 0.000 | 0.000 | 0.000 | 0.010 | 0.010 | 0.0100 | 0.010 |
| 11 | 0.450 | 0.000 | -0.090 | 0.360 | 0.450 | 0.010 | 0.010 | 0.0100 | 0.010 |
| 12 | 0.000 | 0.000 | 0.000 | 0.000 | 0.000 | 0.000 | 0.000 | 0.000 | 0.000 |
| 13 | -0.115 | -0.006 | -0.028 | -0.143 | -0.121 | 0.000 | 0.000 | 0.000 | 0.000 |
| 14 | 0.000 | 0.000 | 0.000 | 0.000 | 0.000 | 0.000 | 0.000 | 0.000 | 0.000 |
| 15 | -0.032 | -0.002 | -0.008 | -0.040 | -0.034 | 0.000 | 0.000 | 0.000 | 0.000 |

|  | Systematic | | | | | Uncertainty | | Random | |
| --- | --- | --- | --- | --- | --- | --- | --- | --- | --- |
| **Skew** | **Geometric** | **Saturation** | **Persistent** | **Injection** | **High field** | **Injection** | **High field** | **Injection** | **High field** |
| 2 | 0.000 | -0.261 | 0.000 | 0.000 | -0.261 | 1.660 | 1.660 | 1.660 | 1.660 |
| 3 | -0.130 | 0.050 | 0.000 | -0.130 | -0.080 | 1.000 | 1.000 | 1.000 | 1.000 |
| 4 | 0.000 | -0.010 | 0.000 | 0.000 | -0.010 | 0.640 | 0.640 | 0.640 | 0.640 |
| 5 | 0.080 | 0.000 | 0.000 | 0.080 | 0.080 | 0.380 | 0.380 | 0.380 | 0.380 |
| 6 | 0.000 | 0.000 | 0.000 | 0.000 | 0.000 | 0.200 | 0.200 | 0.200 | 0.200 |
| 7 | 0.030 | 0.000 | 0.000 | 0.030 | 0.030 | 0.090 | 0.090 | 0.090 | 0.090 |
| 8 | 0.000 | 0.000 | 0.000 | 0.000 | 0.000 | 0.050 | 0.050 | 0.050 | 0.050 |
| 9 | 0.000 | 0.000 | 0.000 | 0.000 | 0.000 | 0.030 | 0.030 | 0.030 | 0.030 |
| 10 | 0.000 | 0.000 | 0.000 | 0.000 | 0.000 | 0.020 | 0.020 | 0.020 | 0.020 |
| 11 | 0.000 | 0.000 | 0.000 | 0.000 | 0.000 | 0.010 | 0.010 | 0.010 | 0.010 |
| 12 | 0.000 | 0.000 | 0.000 | 0.000 | 0.000 | 0.000 | 0.000 | 0.000 | 0.000 |
| 13 | 0.000 | 0.000 | 0.000 | 0.000 | 0.000 | 0.000 | 0.000 | 0.000 | 0.000 |
| 14 | 0.000 | -0.000 | 0.000 | 0.000 | -0.000 | 0.000 | 0.000 | 0.000 | 0.000 |
| 15 | 0.000 | 0.000 | 0.000 | 0.000 | 0.000 | 0.000 | 0.000 | 0.000 | 0.000 |

Table 11-3: Main parameters of the MBH

| Characteristics | Unit | Value |
|---|---|---|
| Aperture | [mm] | 60 |
| Number of apertures | - | 2 |
| Distance between apertures at room temperature/1.9 K | [mm] | 194.52/194.00 |
| Cold mass outer diameter | [mm] | 580 |
| Magnetic length | [m] | 5.307 |
| Coil physical length, as per magnetic design | [m] | 5.415 |
| Magnet physical length: active part (between the end plates) | [m] | 5.799 |
| Magnet physical length: cold mass (between datum planes C and L) | [m] | 6.252 |
| Cold mass weight | [tonne] | ~8 |
| Nominal operation current | [kA] | 11.85 |
| Bore field at nominal current | [T] | 11.23 |
| Peak field at nominal current (without strand self-field correction) | [T] | 11.59 |
| Operating temperature | [K] | 1.9 |
| Load line margin | (%) | 19 |
| Stored energy/m at $I_{nom}$ | [MJ/m] | 0.9663 |
| Differential inductance/m at $I_{nom}$ | [mH/m] | 11.97 |
| Number of layers | - | 2 |
| Number of turns (inner/outer layer) | - | 56 (22/34) |
| Superconductor | - | $Nb_3Sn$ |
| Cable bare width before reaction | [mm] | 14.7 |
| Cable bare mid-thickness before reaction | [mm] | 1.25 |
| Keystone angle | [degree] | 0.79 |
| Cable unit length for the two layers (no layer jump splice) | [m] | ~600 |
| Strand diameter | [mm] | 0.700 ± 0.003 |
| Number of strands per cable | - | 40 |
| Cu to non-Cu ratio | - | 1.15 ± 0.10 |
| RRR, after reaction | - | >100 |
| Minimum strand critical current, $I_c$, without self-field correction (12 T, 4.222 K) | [A] | 438 |
| Minimum strand current density, $J_c$, at 12 T, 4.222 K | [A/mm$^2$] | 2560 |
| Cable insulation thickness per side azimuthal, before/after reaction | [mm] | 0.155/0.110 |
| Heat exchanger hole diameter | [mm] | 60 |
| Heat exchanger distance from centre (same position as in the MB) | [mm] | 180 |
| Cold bore tube inner diameter/thickness (assuming the current CBT is used) | [mm] | 50/1.5 |
| Gap CBT to coil (assuming the current CBT and ground insulation are used) | [mm] | 3 |

### 11.3.3 Protection

The MBH will be protected with quench heaters and a bypass diode operating at cold, integrated with the cold mass assembly. At the current stage of development, it is foreseen to use one bypass diode for the two MBHs of a cryo-assembly. However, this needs to be validated.

### 11.3.4 Radiation

The MBH will, inevitably, see a shower of particles from the collimator. The worst case currently is with ion operation at IP2, for which the peak dose in the coils is estimated to be around 1 MGy [5, 6]. The 11 T dipole cold mass will be designed with a reasonable margin, 5 MGy.

### 11.3.5 Installation and dismantling

The cold mass of the MBH will be equipped with standard features in the ends facing the existing MBs in the tunnel to facilitate the installation and connection, e.g. M-flanges and bellows, preparation of the busbar extremities with regard to splicing, end flanges on the X/V lines, etc. The ends facing the collimator need to be specific; however, standard elements will be used as much as possible.

## 11.4 Inventory of units to be installed and spare policy

Two full-length MBH prototypes will be fabricated to validate the design, the overall performance in nominal operation conditions (to be checked on horizontal test benches in building SM18 at CERN), and the different interfaces with the neighbour systems.

The cryo-assemblies to be installed in LHC are listed below.

- For LS2, around IP2, two cryo-assemblies to replace two main dipoles MB.A10L2 and MB.A10R2, i.e. 4 MBHs and two bypass cryostats.
- For LS3, around IP7, four cryo-assemblies to replace four main dipoles MB.B8L7, MB.B10L7, MB.B8R7, and MB.B10R7, i.e. eight MBHs and four bypass cryostats.
- For LS3, to be confirmed, around IP1 and IP5, up to a maximum of eight cryo-assemblies to replace eight main dipoles, location to be defined.

It is planned to fabricate at least two spare MBHs and one spare bypass cryostat. The prototypes may be used as spares, should they conform fully to the functional requirements.